\documentclass[aps, pra, twocolumn, reprint, nofootinbib, longbibliography]{revtex4-1}

\usepackage{lmodern}
\usepackage{sistyle}
\usepackage{graphicx}
\usepackage{amsmath}
\usepackage{amssymb}
\usepackage{upgreek}
\usepackage{mathrsfs}
\usepackage{verbatim}   % useful for program listings
\usepackage{subfigure}  % use for side-by-side figures
\usepackage{hyperref}
\usepackage{xcolor}
\hypersetup{
	colorlinks,
	linkcolor={red!95!black},
	citecolor={blue!100!black},
	urlcolor={green!45!black}
}
\usepackage{verbatimbox}

\newcommand{\um}{\upmu\text{m}}

\newcommand{\uC}{\upmu\text{C}}
\newcommand{\SiN}{Si$_3$N$_4$~}

\newcommand{\fig}[1]{~\ref{fig#1}\hyperref[#1]}

\begin{document}

\title
  {Etch-Tuning and Design of Silicon Nitride Photonic Crystal Reflectors}

\author{Simon Bernard,$^1$ Christoph Reinhardt,$^1$ Vincent Dumont,$^1$ Yves-Alain Peter,$^2$ Jack C. Sankey$^1$}
\affiliation{$^1$Department of Physics, McGill University, Montr\'{e}al, 
	Qu\'{e}bec, H3A 2T8, Canada}
\affiliation{$^2$Department of Engineering Physics, Polytechnique, Montr\'{e}al, 
	Qu\'{e}bec, H3C 3A7, Canada}

\email{jack.sankey@mcgill.ca}


\begin{abstract}
	By patterning a freestanding dielectric membrane into a photonic crystal reflector (PCR), it is possible to resonantly enhance its normal-incidence reflectivity, thereby realizing a thin, single-material mirror. In many PCR applications, the operating wavelength (e.g.~that of a low-noise laser or emitter) is not tunable, imposing tolerances on crystal geometry that are not reliably achieved with standard nanolithography. Here we present a gentle technique to finely tune the resonant wavelength of a \SiN PCR using iterative hydrofluoric acid etches. With little optimization, we achieve a 57-nm-thick photonic crystal having an operating wavelength within 0.15 nm (0.04 resonance linewidths) of our target (1550 nm). Our thin structure exhibits a broader and less pronounced transmission dip than is predicted by plane wave simulations, and we identify two effects leading to these discrepancies, both related to the divergence angle of a collimated laser beam. To overcome this limitation in future devices, we distill a series of simulations into a set of general design considerations for realizing robust, high-reflectivity resonances.
\end{abstract}

% possibly not necessary for ACS mode
\maketitle


% flat lenses \cite{Lu2010Planar,Fattal2010Flat},
Two-dimensional (2D) photonic crystals provide a powerful means of controlling the flow of light \cite{Joannopoulos1995Photonic}, and can be engineered so that ``leaky'' guided modes produce strong, resonant interactions with waves propagating out of the crystal plane \cite{Zhou2014Progress}. On resonance, these structures can be highly reflective, enabling (among other things) compact etalon filters \cite{Suh2003Displacement, Ho2015Two}, lightweight laser mirrors \cite{Huang2007A, Yang2015Laser}, nonlinear optical elements \cite{Yacomotti2006All} and biochemical sensors \cite{Shi2009Tunable, Magnusson2011Resonant}. Free-standing photonic crystal reflectors (PCRs) \cite{Fan2002Analysis, Lousse2004Angular, Kilic2004Photonic} represent a particularly advantageous technology in the field of optomechanics \cite{Aspelmeyer2014Cavity}, wherein it is desirable to create a lightweight mass that strongly reflects incident light, thereby maximizing the influence of each photon. With this aim, single-layer reflectors have been fabricated from InP \cite{Antoni2011Deformable, Antoni2012Nonlinear, Makles20152DARXIV, Yang2015Laser} and SiN \cite{Bui2012High, Kemiktarak2012Cavity, Kemiktarak2012Mechanically, Stambaugh2014From, Norte2016Mechanical, Chen2016HighARXIV}. High-stress \SiN is of particular interest, since it combines ease of fabrication, low optical loss \cite{Wilson2009Cavity, Sankey2010Strong}, and the lowest mechanical dissipation and force noise \cite{Reinhardt2016Ultralow, Norte2016Mechanical}. Furthermore, incorporating a PCR does not introduce significant mechanical losses \cite{Bui2012High, Norte2016Mechanical}.

An outstanding goal is to fabricate PCRs with the high-reflectivity resonance at a specific design wavelength (e.g., to match that of a low-noise laser, an atomic transition, or another reflector). However, the width of a typical 2D SiN PCR resonance is measured in nanometers, imposing geometrical tolerances that cannot be reliably achieved with standard nanolithography. Inspired by similar work on microdisks resonators and photonic crystal defect cavities in other materials \cite{Barclay2006IntegrationARXIV, White2005Tuning, Henze2013Fine, Dalacu2005Postfabrication}, we present a fabrication technique capable of achieving the desired precision through iterative tuning of the resonance with hydrofluoric (HF) acid dips. In this proof-of-concept study, we systematically tune the resonance of a 60-nm-thick photonic crystal by 16 nm, such that it resides within 0.15 nm (0.04 linewidths) of our targeted wavelength $\lambda_0$=$1550$ nm. This is achieved with a 2D square lattice of holes -- chosen to eliminate any polarization-dependent response -- but should be readily applicable to 1D gratings \cite{Kemiktarak2012Cavity, Kemiktarak2012Mechanically, Stambaugh2014From} and other related structures.

A second question is whether there exists a fundamental and/or practical limitation on how thin (lightweight) these structures can be made while still maintaining high reflectivity. For example, despite our comparatively low feature roughness, our thin structures exhibits a resonant dip in the transmission spectrum that is broader and not as deep as predicted by simple plane-wave simulations, reaching a minimal value of only 0.32. This discrepancy has been observed in similar crystals of thickness at or below 100 nm \cite{Bui2012High, Norte2016Mechanical}, and has been attributed to fabricated disorder not included in periodic-crystal simulations \cite{Bui2012High}. Since then, however, thicker square-lattice PCRs -- having comparable disorder to our own -- achieve a transmission dip well below 1\% \cite{Norte2016Mechanical, Chen2016HighARXIV}. Building upon previous investigations of angular- and collimation-induced changes in the transmission spectrum of optically thick structures \cite{Crozier2006Air}, we present simulations illustrating two effects leading to collimation-induced resonance broadening: (i) the previously identified angular dependence of the ``primary'' resonance wavelength \cite{Lousse2004Angular,Crozier2006Air}, and (ii) ``parasitic'' crystal modes that couple only to off-normal plane waves \cite{Lousse2004Angular,Crozier2006Air,Bui2012High,Chen2016HighARXIV}, which can strongly interfere with the primary resonance. Finally, we perform a series of simulations for crystals of varied geometry, and distill the results into a guide for reliably minimizing (or balancing) these effects to realize a maximally robust high-reflectivity resonance.

\section{Etch-Tuning the Resonance}\label{sec:exp}

Our primary goal is to tune the resonance of a thin \SiN PCR to a convenient (telecom) laser wavelength $\lambda_0$=\SI{1550}{nm}. Since electron beam lithography cannot achieve the required tolerances on its own, we first fabricate a structure intentionally having too much material, then iteratively measure the transmission spectrum $\mathcal{T}$ and etch with HF until the resonant wavelength $\lambda_r$ (here defined at the minimal value of $\mathcal{T}$) is close to $\lambda_0$. 

\begin{figure}
	\centering
	\includegraphics[width=0.9\columnwidth]{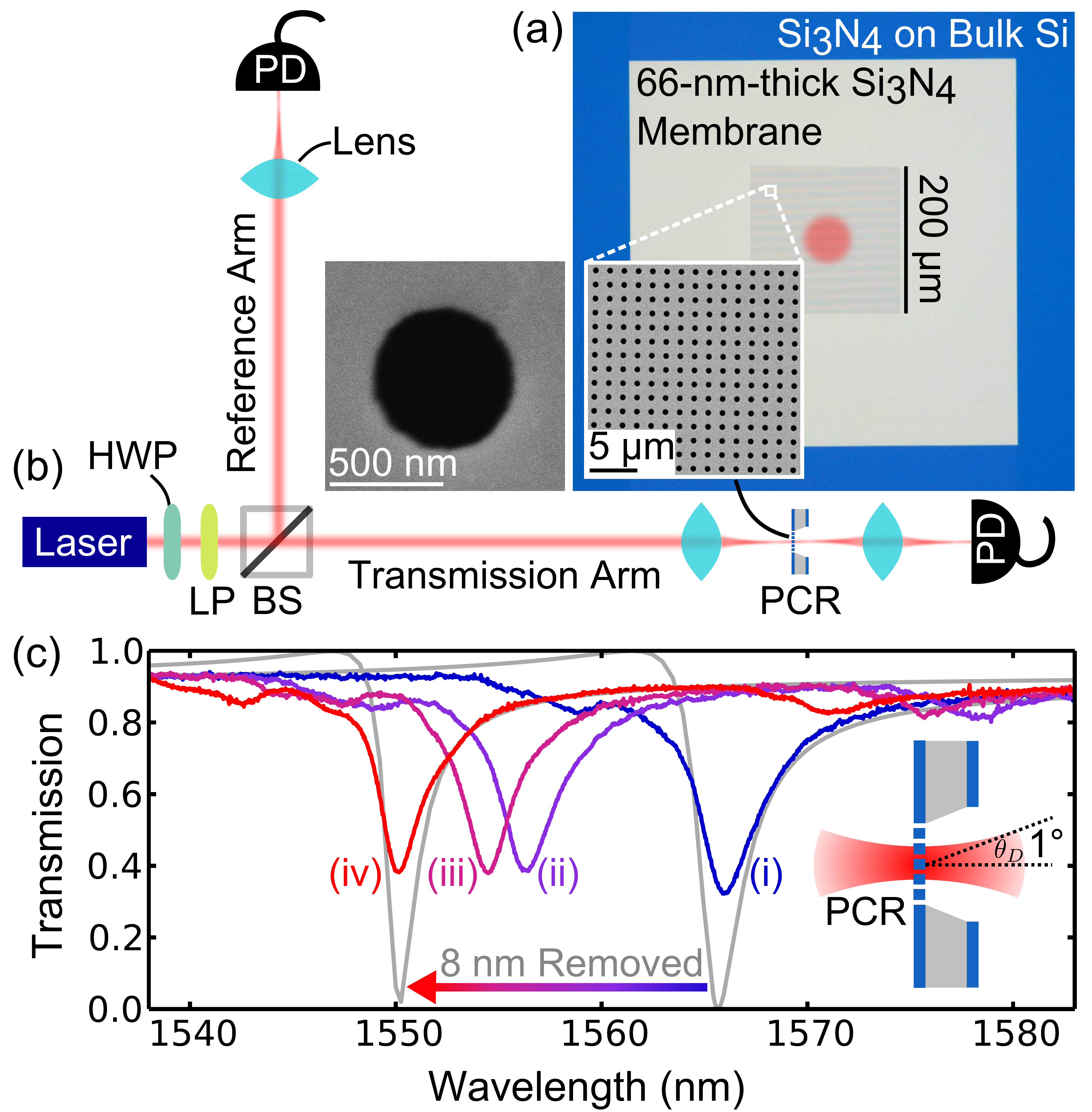}
	\caption{Etch-tuning the resonance. (a) \SiN free-standing photonic crystal reflector. Inset SEM images show the initially fabricated PCR, with thickness $h$=$66 \pm 1$ nm, hole diameter $d$=$614$ nm ($\pm$7 nm hole-to-hole with edge roughness $\sim 15$ nm) and lattice constant $a$=$1500 \pm 6$ nm. The red spot indicates the approximate $1/e^2$ diameter $D$=$60 \pm 1~\um$ of the laser (divergence angle $\theta_D\sim1^\circ$ drawn in (c)). (b) Transmission measurement: laser beam passes through a half-wave plate (HWP), linear polarizer (LP), beam splitter (BS), focusing lens, and the crystal, and is collected by a photodiode (PD) to estimate the transmitted power. Power fluctuations are monitored with a second ``reference arm'' PD. (c) Transmission spectrum $\mathcal{T}$ of the crystal (i) as fabricated and then iteratively immersed in HF for (ii) 130 s, and (iii) 165 s and (iv) 195 s (total), corresponding to an integrated thickness change of $9 \pm 1$ nm (measured by a reflectometer) and hole diameter increase of $\sim$ 2 nm (estimated from SEM images). Gray curves show the simulated response (MEEP \cite{Oskooi2010Meep}) for normal-incidence plane waves, assuming a \SiN refractive index of 2, hole diameters (i) 614 nm and (iv) 618 nm and thicknesses (i) 62.5 nm and (iv) 54.5 nm.}
	\label{fig1}
\end{figure}

The crystal is initially fabricated from a commercial Si wafer coated on both sides with 100-nm-thick stoichiometric \SiN deposited by low-pressure chemical vapor deposition (LPCVD, University Wafers). The unpatterned membrane is defined by opening a square window in the back-side nitride using optical lithography and reactive ion etching (RIE), then etching through the exposed silicon in a $45\%$ KOH solution at $75\,^{\circ}$C ($\sim$ 35 $\um/$hr etch rate) to release the front-side nitride membrane. The top surface of the nitride is then coated with a 150-nm-thick resist (Zeon Chemicals ZEP520A diluted in anisole with a 1:2 ratio, spun at 3000 rpm and baked for 40 min in an oven at $180\,^{\circ}$C), and exposed in an electron-beam writer (50 $\uC/\text{cm}^2$ dose at 10 kV) to define the crystal's etch mask. The resist is developed for 60 s (Zeon Chemicals ZED-N50), and rinsed for 15 s in isopropyl alcohol. The pattern is then transfered to the membrane via RIE in a mixture of CHF$_3$ (30 sccm), CF$_4$ (70 sccm), and Ar (7 sccm) with a 30 mTorr chamber pressure and 100 W power. The remaining resist is stripped with Microposit Remover 1165 for 30 min at $60\,^{\circ}$C and a 3:1 (H$_2$SO$_4$:H$_2$O$_2$) piranha solution for 15 min. Finally, the sample is cleaned and partially etched for 10 min in 10:1 HF acid at room temperature,\footnote{Note we generally do not etch through the full nitride on the RIE step, allowing the final HF dip to finish the job. This reduces crystal breakage during the RIE step. We also use a carrier that holds chips rigidly and vertically in solution, as in Ref.~\cite{Reinhardt2016Ultralow} (designs available upon request).} then rinsed with water and methanol. This initial structure has nominal thickness $h$=$66\pm1$ nm (measured by a reflectometer) and holes of diameter $d $=$ 614$ nm ($\pm$7 nm hole-to-hole with edge roughness $\sim 15$ nm), in a square lattice spaced by $a $=$ 1500 \pm 6$ nm (Fig.\fig{1}(a)).

We can locate the crystal resonance with the swept-laser transmission measurement sketched in Fig.\fig{1}(b). A collimated beam from a fiber-coupled tunable laser (New Focus Velocity 6328) is focused to a Gaussian spot having $1/e^2$ intensity diameter $D $=$ 60 \pm 1$ $\um$ at the crystal (relative spot size indicated by a red circle in Fig.\fig{1}(a)) and collected by a photodiode (PD). Fluctuations in the incident power are monitored by a second photodiode (``reference arm''). A rotatable half-wave plate (HWP) and linear polarizer (LP) define the input polarization to avoid any polarization-dependence of the beam splitter (BS) and other optics (the PCR response is found to be polarization-independent, however). Figure\fig{1}(c) shows transmission $\mathcal{T}$ versus wavelength $\lambda$ for the crystal (i) as fabricated and after (ii) 130 s, (iii) 165 s, and (iv) 195 s of HF etching (total). Immersing the structure in HF simultaneously decreases its thickness (at a rate $\sim$ 3 nm/min) and increases the hole diameters (rate $\sim$ 1  nm/min), displacing the resonance toward shorter wavelengths. Using this technique, the location of the resonance is tuned within 0.15 nm of the target wavelength (red curve). Here the shift in $\lambda_r$ per minute in HF is found to fluctuate somewhat ($-4.4$ nm/min (i $\rightarrow$ ii), $-3.2$ nm/min (ii $\rightarrow$ iii), and $-7.9$ nm/min (iii $\rightarrow$ iv)), which can be mitigated by using a slower, buffered solution and introducing gentle fluid circulation.

The gray curves in Fig.\fig{1}(c) show the simulated response for a normal-incidence infinite plane wave. The hole diameters, which have a comparatively weak effect on the resonant wavelength $\lambda_r$, are set to (i) 614 nm and (iv) 618 nm to match the values extracted from SEM images (e.g.~Fig.\fig{1}(a,inset)), and the thicknesses are set to (i) 62.5 nm and (iv) 54.5 nm to match the observed $\lambda_r$. These thickness values are within 6\% of the those measured independently by a reflectometer, and the simulated thickness change of 8 nm is consistent with the measured change of $9 \pm 1$ nm. 

The linewidth of the measured resonance is also reduced by 20\% as the device is thinned, qualitatively consistent with a 30\% decrease predicted by the simulations, though the measured resonance is also a factor of $\sim$ 1.6 broader than the simulations suggest. Moreover, the depth of the resonance reaches a minimum value $\mathcal{T}_\text{min}$ between 0.32 ($h$=66 nm) and 0.38 ($h$=57 nm), placing an upper bound of 0.68 ($h$=66 nm) and 0.62 ($h$=57 nm) on the reflectivity. As discussed below, the vast majority of the resonance broadening in this case (and other thin crystals) likely arises from $\mathcal{T}$'s sensitivity to incident angle and the superposition of incident plane waves present in a collimated beam. 

\section{Collimation-Induced Broadening}\label{sec:design}

A practical limitation in the performance of photonic crystals is the sensitivity of the resonance to the radiation's angle of incidence \cite{Lousse2004Angular,Crozier2006Air}: since collimated beams comprise a weighted superposition of plane waves from all angles $\theta$ (e.g.~spanning $\theta_D \sim 1^\circ$ for our $60$-$\um$ spot; see inset of Fig.\fig{1}(c)), this sensitivity can average away the effect of crystal resonances. One can compensate by increasing $D$ to reduce $\theta_D$, but this in turn requires a larger-area crystal, which is disadvantageous for many applications, and furthermore difficult to fabricate: the crystal in Fig.\fig{1}~uses the full $200\times 200$ $\um^2$ high-resolution write field of our electron-beam writer -- using a larger field reduces precision, and using multiple fields introduces detrimental dislocations between adjacent fields. Our beam diameter of 60 $\um$ was chosen to be large but still safely contained within the high-resolution field of our electron-beam writer.

\begin{figure}
	\centering
	\includegraphics[width=0.9\columnwidth]{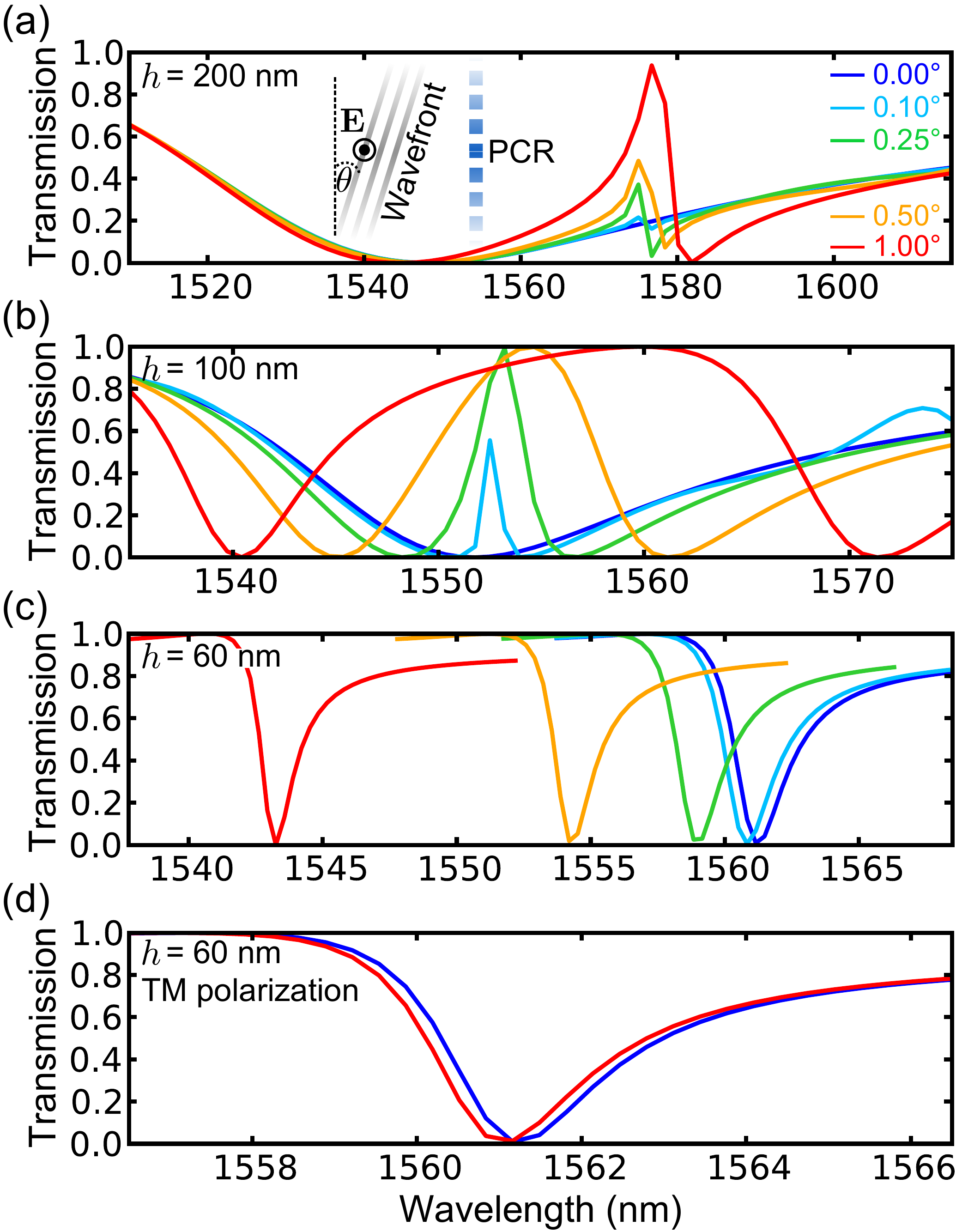}
	\caption{Simulated transmission spectrum for plane waves of incident angles $\theta$=0 (blue), 0.1$^\circ$ (cyan), 0.25$^\circ$ (green), 0.5$^\circ$ (orange), and 1$^\circ$ (red) passing through (a) $h$=$200$-nm-thick crystal with $a$=$1339$ nm, $d$=$1044$ nm, (b) $h$=$100$-nm-thick crystal with $a$=$1470$ nm, $d$=$1109$ nm, and (c), (d) a $h$=$60$-nm-thick crystal with $a$=$1500$ nm, $d$=$616$ nm. The inset shows the incident angle and TE polarization used for (a), (b), and (c). The effect of a nearly degenerate ``parasitic'' mode can be seen in (b), and (c) shows the tilt-dependence of the resonant wavelength $\lambda_r \propto \theta^2$. (d) The TM polarization exhibits comparatively little angular dependence.}
	\label{fig2}
\end{figure}

To get a sense of when this broadening is important, Fig.\fig{2}(a) shows a set of illustrative simulations for thin \SiN crystals of varied geometry and plane wave incidence angles $\theta$ between 0$^\circ$ and 1$^\circ$. The transverse-electric (TE) polarization (inset) is chosen in (a)-(c) to maximize the effect of $\theta$ on $\mathcal{T}$; for the transverse-magnetic (TM) polarization, $\mathcal{T}$ is comparatively unaffected (see (d)). The 200-nm-thick crystal in (a) exhibits a single, broad ``primary'' resonance near 1550 nm for $\theta$=$0$ (blue curve), and as $\theta$ is increased, a second ``parasitic'' resonance becomes increasingly pronounced. The existence of these additional resonances is well-known: they do not appear in normal-incidence transmission spectra because symmetry precludes coupling to such modes \cite{Fan2002Analysis, Crozier2006Air, Bui2012High, Chen2016HighARXIV}. For $\theta\ne 0$, however, its presence has a modest impact on the spectrum, shifting the minimum in $\mathcal{T}$ to lower $\lambda$. Figure\fig{2}(b) shows the angular dependence for a 100-nm-thick crystal. In this case, the separation $\Delta$ between the primary and parasitic resonance is only a few nanometers, and the effect is profound over the full range of angles. One would not expect high reflectivity for any reasonably collimated beam. Figure\fig{2}(c) illustrates another cause of collimation-induced broadening \cite{Crozier2006Air} that can occur in geometries where $\Delta$ is large: $\lambda_r$ in general varies quadratically with $\theta$. The geometry of Fig.\fig{2}(c) is close to that of our devices in Fig.\fig{1}, implying this is likely the dominant mechanism in our system. Broadening has been observed at normal incidence for similarly thin \SiN crystals \cite{Bui2012High, Norte2016Mechanical}, and in Ref.~\cite{Bui2012High}, the resonance furthermore was found to split, broaden, and shallow when tilting the crystal away from normal incidence \cite{Bui2012High}, providing evidence for the influence of a second mode. 
 
\section{Design Considerations}\label{sec:disc}

\begin{figure} % had to reduce width from 0.9 to 0.87
	\centering
	\includegraphics[width=0.80\columnwidth]{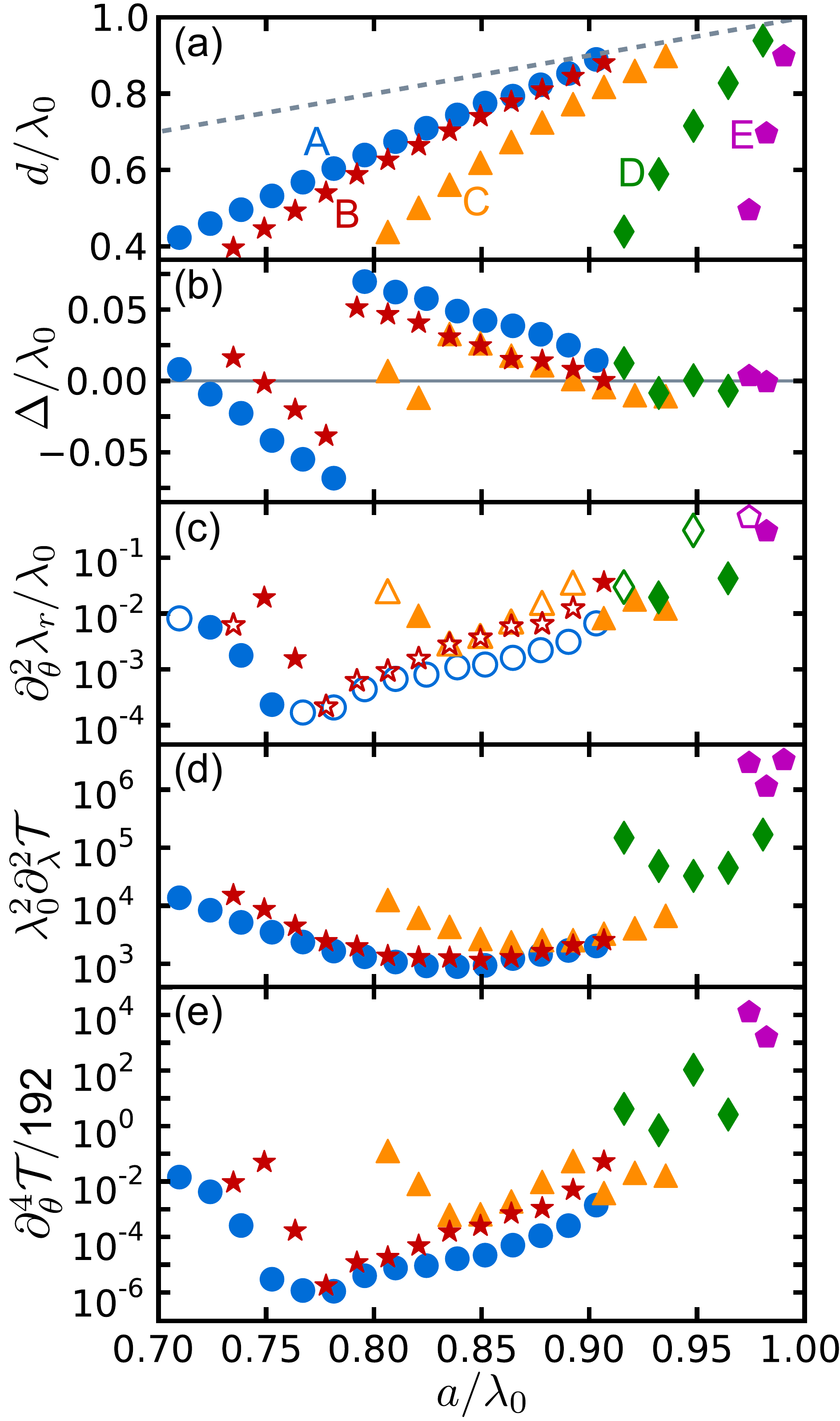}
	\caption{Parameter space of \SiN PCRs (scaled by the targeted wavelength $\lambda_0$). (a) Lattice constants and hole diameters producing a resonant wavelength $\lambda_r$ within 0.3\% of $\lambda_0$ for thicknesses $h$=$0.26 \lambda_0$ (circles; $h$=$400$ nm for $\lambda_0$=$1550$ nm), $h$=$0.19 \lambda_0$ (stars, 300 nm), $h$=$0.13 \lambda_0$ (triangles, 200 nm), $h$=$0.065 \lambda_0$ (diamonds, 100 nm), and $h$=$0.032 \,\lambda_0$ (pentagons, 50 nm). The dashed line indicates the limiting geometry $d $=$ a$. (c) Quadratic angular sensitivity $\partial_\theta^2\lambda_r$ ($\theta$ measured in degrees). Hollow symbols denote a negative value. (d) Curvature $\partial_\lambda^2\mathcal{T}$ of the normal-incidence transmission spectrum $\mathcal{T}$ at $\lambda_r$. (e) Figure of merit for collimation broadening. A, B, C, D and E denote $(a,d)$ combinations that minimize the figure of merit. Note the $h$=100 and 50 nm sets (b)-(d) are missing a data point due to numerical convergence issues.}
	\label{fig3}
\end{figure}

This motivates two basic PCR design goals: maximize the separation $\Delta$ from the parasitic resonances, and minimize the resonant wavelength's sensitivity $\partial_\theta^2 \lambda_r$ to incidence angle $\theta$. Figure\fig{3}~shows a summary of simulations for varied thickness $h$, hole diameter $d$, and lattice constant $a$, chosen so that the resonance wavelength $\lambda_r$ remains near a ``design'' wavelength $\lambda_0$. All values are scaled by $\lambda_0$ to facilitate application to other systems having a dielectric refractive index $n$ near $2$. For each thickness, the desired resonance can be achieved over a wide range of ($a$,$d$) combinations, as shown in (a). Figure\fig{3}(b) shows the detuning $\Delta$ of the nearest parasitic resonance, which can be made largest in the thickest devices (especially near $a$=$0.78\lambda_0$). 

The angular sensitivity $\partial_\theta^2 \lambda_r$ plotted in (c) shows a general tendency toward lower values for thicker devices, but is strongly influenced by parasitic modes whenever $\Delta$ approaches 0, leading to complex behavior. Interestingly, the sign of $\partial_\theta^2 \lambda_r$ often flips as a function of $a$, implying the existence of an intermediate geometry in which all of these effects balance, thereby eliminating collimation broadening to lowest order. The three thickest devices exhibit a well-resolved global minimum in $\partial_\theta^2\lambda_r$ near a sign flip, as expected. Presumably the other geometries having sign flips also exhibit such minima, but they are not resolved at this level. This will remain the subject of future study, but we will note three caveats at the outset: (i) $\partial_\theta^2 \lambda_r$ is particularly sensitive to the geometry near these points (especially in thin devices), which tightens their fabrication requirements, (ii) it will be challenging to precisely tune both $\Delta$ and $\lambda_r$ with existing fabrication techniques (including HF etching), and (iii) we expect higher-order $\theta$-dependencies to play a role for any reasonably collimated beam (even $\theta_D\sim 1^\circ$), and especially in the thinnest devices where $\partial_\theta^2 \lambda_r$ is largest. 

While a small value of $\partial_\theta^2 \lambda_r$ is certainly desirable, this quantity alone is not a useful figure of merit for achieving high reflectivity, since the impact on narrower resonances -- characterized by a larger curvature $\partial_\lambda^2\mathcal{T}$ at $\lambda_r$; see Fig.~\ref{fig3}(d) -- is larger than for wider resonances. With this in mind, we note that, to lowest order, $\mathcal{T} \approx \frac{1}{2}(\partial^2_\lambda\mathcal{T})(\lambda-\lambda_r)^2$ near resonance, where $\lambda_r$ varies with $\theta$ as $\approx\frac{1}{2}(\partial^2_\theta \lambda_r)\theta^2$. Averaging this approximate $\mathcal{T}$ over all incident angles of the collimated beam\footnote{Note this treatment (inspired by the numerical approach in Ref.~\cite{Crozier2006Air}) is only valid for an infinite crystal and small $\theta_D$, such that the transmitted field from each incident plane wave comprises a discrete set of outbound plane waves that are orthogonal to those produced by the other incidence waves.} (i.e.~weighted by the Gaussian distribution $\propto e^{-2\theta^2/\theta_D^2}$), produces a shift in the minimum's wavelength $\delta\lambda_{r}\approx\frac{1}{8}\left(\partial_{\theta}^{2}\lambda_{r}\right)\theta_{D}^{2}$ and a new minimum value
\begin{eqnarray}
\mathcal{T}_\text{min}&\approx&\frac{1}{64}\left(\partial_{\lambda}^{2}\mathcal{T}\right)\left(\partial_{\theta}^{2}\lambda_{r}\right)^{2}\theta_{D}^{4}\\&=&\frac{1}{192}\left(\partial_{\theta}^{4}\mathcal{T}\right)\theta_{d}^{4}.
\end{eqnarray}
A more reasonable figure of merit for these structures is therefore the prefactor $\frac{1}{192}\partial_\theta^4\mathcal{T}$ (the approximate value of which is plotted in Fig.~\ref{fig3}(e)), which can be used to directly compare the relative performance of different geometries, and to roughly estimate $\mathcal{T}_\text{min}$ for sufficiently small $\theta_D$ (i.e.~provided $\mathcal{T}_\text{min} \ll 1$). The thickest crystals simulated here (crystal ``A'' having $a\sim 0.78\lambda_0$) should be the most robust against these effects, with a collimation-limited $\mathcal{T}_\text{min} \lesssim 10^{-6}$ for a $\theta_D$=$1^\circ$ beam. Additionally, the broader resonances (Fig.~\ref{fig3}(d)) and $a$-dependence of $\partial_\theta^4\mathcal{T}$ (Fig.~\ref{fig3}(e)) will further relax fabrication tolerances, while the added thickness reduces breakage during release (these and thicker structures may require low-stress nitride to fabricate, however, due to the stress-induced cracking in thick layers of LPCVD nitride on Si). Note that these plotted values represent approximate lowest-order dependences, and real devices will exhibit other nonidealities not included here: disorder, scattering, and absorption. 

Finally, we estimate the etch tuning rate $\partial_h \lambda_r$ near the ``optimal'' crystal parameters labeled A, B, C, D, and E in (a), finding values $\sim$ 0.6, 1.0, 1.3, 1.5, and 0.9, respectively. This quantity does not vary wildly with thickness, but it does suggest that etch tuning will also be marginally more controlled for the thickest optimal device. Away from these points, $\partial_h \lambda_r$ similarly varies by $\sim$ 50\%.

\begin{table} % may have to use [H] after the text is final
	\begin{center}%\addvbuffer[14pt 0pt]{
			\begin{tabular}{ c | c | c | c | c | c | c | c}
				\hline
				Ref. & $\lambda_r$ (nm) & $h/\lambda_r$ & $a/\lambda_r$ & $d/\lambda_r$ & $\mathcal{T}_\text{min}$ (\%) & $\theta_D$ ($^\circ$) & $n$ \\ \hline
				\cite{Crozier2006Air} & 763 & 0.26 & 0.59 & 0.25 & 6 & 2 & 2.3 \\
				\cite{Crozier2006Air} & 763 & 0.26 & 0.59 & 0.25 & 15 & 3 & 2.3 \\
				\cite{Chen2016HighARXIV} & 1071 & 0.19 & 0.78 & 0.55 & 0.053 & 0.41 & 2 \\
				\cite{Norte2016Mechanical} & 1546 & 0.13 & 0.88* & 0.72* & 0.7 & 1.1* & 2 \\
				\cite{Norte2016Mechanical} & 1547 & 0.097 & 0.91* & 0.69* & 2.8 & 1.1* & 2 \\
				\cite{Norte2016Mechanical} & 1517 & 0.066 & 0.97* & 0.66* & 17 & 1.1* & 2 \\
				\cite{Bui2012High} & 1035 & 0.048 & 0.93 & 0.28 & 43 & - & 2.2 \\
				This Work & 1566    & 0.042 & 0.96 & 0.39 & 32 & 0.94 & 2 \\
				This Work & 1550.14 & 0.037 & 0.97 & 0.40 & 38 & 0.94 & 2 \\
				\hline		
			\end{tabular}
		%}
	\end{center}
	\caption{Summary of previous work on \SiN or SiN$_x$ square lattice PCRs (resonance $\lambda_r$, thickness $h$, lattice constant $a$, hole diameter $d$, minimum transmission $\mathcal{T}_\text{min}$, divergence angle $\theta_D$, and refractive index $n$). Stars (*) indicate values obtained via personal communication.}
	\label{table1}
	
\end{table}

\section{Summary and Discussion}

We have introduced a simple technique for iteratively tuning the resonance wavelength of a photonic crystal reflector, achieving a value within 0.15 nm (0.04 linewidths) of our target. Consistent with literature, we observe higher transmission than is predicted by simple plane wave simulations, and identify two fundamental limitations on reflectivity imposed by collimated light. Table~\ref{table1} lists the parameters reported in previous work on SiN PCRs, which shows some of the expected trends in $\mathcal{T}(\lambda_r)$ with $h$ and $\theta_D$, despite the varied differences in $\lambda_r$, $(a,d)$ combinations, and index (material composition). We observe a $\mathcal{T}(\lambda_r)$ that is somewhat improved over a similarly thin structure (Ref.~\cite{Bui2012High}), but our hole diameter is also significantly larger, resulting in a wider resonance that is less sensitive to $\theta$, and our operating wavelength is $\sim 50$\% longer, relaxing our fabrication tolerances and reducing absorption. A larger deviant from the trend is Ref.~\cite{Crozier2006Air}, but in that case the crystal parameters are quite far from the optimal point identified in Fig.~\fig{3}, and they use a shorter wavelength on non-stoichiometric SiN, both of which lead to higher absorption.

Within the context of optomechanics and force sensing, these results illustrate an important trade-off between mechanical mass and attainable reflectivity. In particular, it is the thinnest \SiN structures (i.e.~well below $100$ nm) \cite{Reinhardt2016Ultralow, Norte2016Mechanical} that exhibit the lowest force noise, leading to an additional trade-off between reflectivity and mechanical performance, or requiring a more complicated structure with regions of different thickness \cite{Norte2016Mechanical}.

\vspace*{0.4cm}

\section*{Acknowledgments}

We thank Alireza H. Mesgar for initial help, Alex Bourassa for data acquisition help, Don Berry, Paul Blond\'{e}, Jun Li, Lino Eugene, and Mattieu Nannini for fabrication help, and Tina M\"{u}ller, Max Ruf, and Abeer Barasheed for fruitful discussions. The group also gratefully acknowledges computational support from CLUMEQ and financial support from NSERC, FRQNT, the Alfred P. Sloan Foundation, CFI, INTRIQ, RQMP, CMC Microsystems, and the Centre for the Physics of Materials at McGill.

\bibliography{Alles}

\end{document}